\documentclass[usenatbib]{mn2e}

\usepackage{dcolumn}
\usepackage{graphics,epsfig}
\usepackage{txfonts}
\usepackage{times}
\usepackage{amssymb}

\newcommand{\dal}{{[\Delta \alpha/ \alpha]}}

\newcommand{\dmu}{{[\Delta \mu/\mu]}}
\newcommand{\lsb}{\left[}
\newcommand{\rsb}{\right]}

\newcommand{\kms}{km~s$^{-1}$}
\newcommand{\cm}{cm$^{-2}$}
 
\newcommand{\NHI}{N_{\rm HI}}
\newcommand{\noi}{\noindent}
\newcommand{\lb}{\left(}

\newcommand{\rb}{\right)}
\newcommand{\pks}{PKS~0201+113}
\newcommand{\ts}{{\rm T_s}}
\newcommand{\beq}{\begin{equation}}
\newcommand{\eeq}{\end{equation}}

\title[HI 21cm absorption at $z \sim 3.39$]{HI 21cm absorption at $z \sim 3.39$ towards PKS~0201+113}
\author[Kanekar et al.]{N.~Kanekar$^1$\thanks{E-mail: nkanekar@aoc.nrao.edu (NK); chengalu@ncra.tifr.res.in (JNC); wendy.peters@nrl.navy.mil (WML)} 
J.~N.~Chengalur$^{2,3}$, W.~M.~Lane$^4$ \\
$^1${}National Radio Astronomy Observatory, 1003 Lopezville Rd, Socorro, NM 87801, USA; 
$^2${}National Centre for Radio Astrophysics, Ganeshkhind, \\ Pune-411007, India;
$^3${}Australia Telescope National Facility, CSIRO, Epping, NSW 1710, Australia; 
$^4${}Naval Research Laboratory, Code 7213, \\ 4555 Overlook 
Ave SW, Washington, DC 20375, USA}

\begin{document}
\date{Received mmddyy/ accepted mmddyy}
\maketitle
\label{firstpage}

\begin{abstract}

We report the Giant Metrewave Radio Telescope detection of HI~21cm absorption 
from the $z \sim 3.39$ damped Lyman-$\alpha$ absorber (DLA) towards 
PKS~0201+113, the highest redshift at which 21cm absorption has been detected 
in a DLA. The absorption is spread over $\sim 115$~km~s$^{-1}$ and has two 
components, at $z = 3.387144 (17)$ and  $z = 3.386141 (45)$. The 
stronger component has a redshift and velocity width in agreement with the 
tentative detection of Briggs et~al.~(1997), but a significantly lower optical depth.  
The core 
size and DLA covering factor are estimated to be $\lesssim 100$~pc and 
$f \sim 0.69$, respectively, from a VLBA 328~MHz image. If one makes
the conventional assumption  that the HI column densities towards the
optical and radio cores are the same, this optical depth corresponds to
a spin temperature of  $\ts \sim [(955 \pm 160) \times (f/0.69)] $~K. However, 
this assumption may not be correct, given that no metal-line absorption 
is seen at the redshift of the stronger 21cm component, indicating that 
this component does not arise along the line of sight to the optical QSO,
and that there is structure in the 21cm absorbing gas on scales smaller 
than the size of the radio core. We model the 21cm absorbing gas as 
having a two-phase structure with cold dense gas randomly distributed 
within a diffuse envelope of warm gas. For such a model, our radio data 
indicate that, even if the optical QSO lies along a line-of-sight with a 
fortuitously high ($\sim 50$\%) cold gas fraction, the average cold gas fraction is low, 
($\lesssim 17\%$), when averaged over the the spatial 
extent of the radio core. Finally, the large mismatch between peak 21cm 
and optical redshifts and the complexity of both profiles makes it unlikely
that the $z \sim 3.39$ DLA will be useful in tests of fundamental constant
evolution.

\end{abstract}


\section{Introduction}
\label{intro}

While high redshift damped Lyman-$\alpha$ absorbers (DLAs), the highest HI column density 
systems seen in quasar spectra, are presumably the precursors of today's galaxies 
and the primary gas reservoir for star formation, their size and structure have long 
been subjects of controversy (e.g. \citealt{wolfe05}).  Understanding the nature 
of a ``typical'' DLA as a function of redshift is one of the important open 
problems in galaxy evolution.

HI 21cm absorption studies of DLAs towards radio-loud quasars provide an estimate 
of the harmonic mean spin temperature of the absorbing gas, via a comparison between 
the 21cm optical depth and the HI column density obtained from the Lyman-$\alpha$ line.
This gives the distribution of gas in different temperature phases, which can be used to 
probe the redshift evolution of physical conditions in the absorbers (e.g. \citealt{kanekar04}). 
Further, a comparison between the redshifts of 21cm and optical metal-line absorption 
(in a statistically large sample) can be used to study the evolution of a 
combination of fundamental constants \citep{wolfe76}. 
Extending 21cm absorption studies  to the highest redshifts is interesting from 
both these perspectives. However, despite  numerous searches, there are still only 
four confirmed detections of 21cm absorption in $z > 1$ DLAs, out to $z \sim 2.347$ 
\citep{wolfe79,wolfe81,wolfe85,kanekar06}.

\begin{figure*}
\centering
\epsfig{file=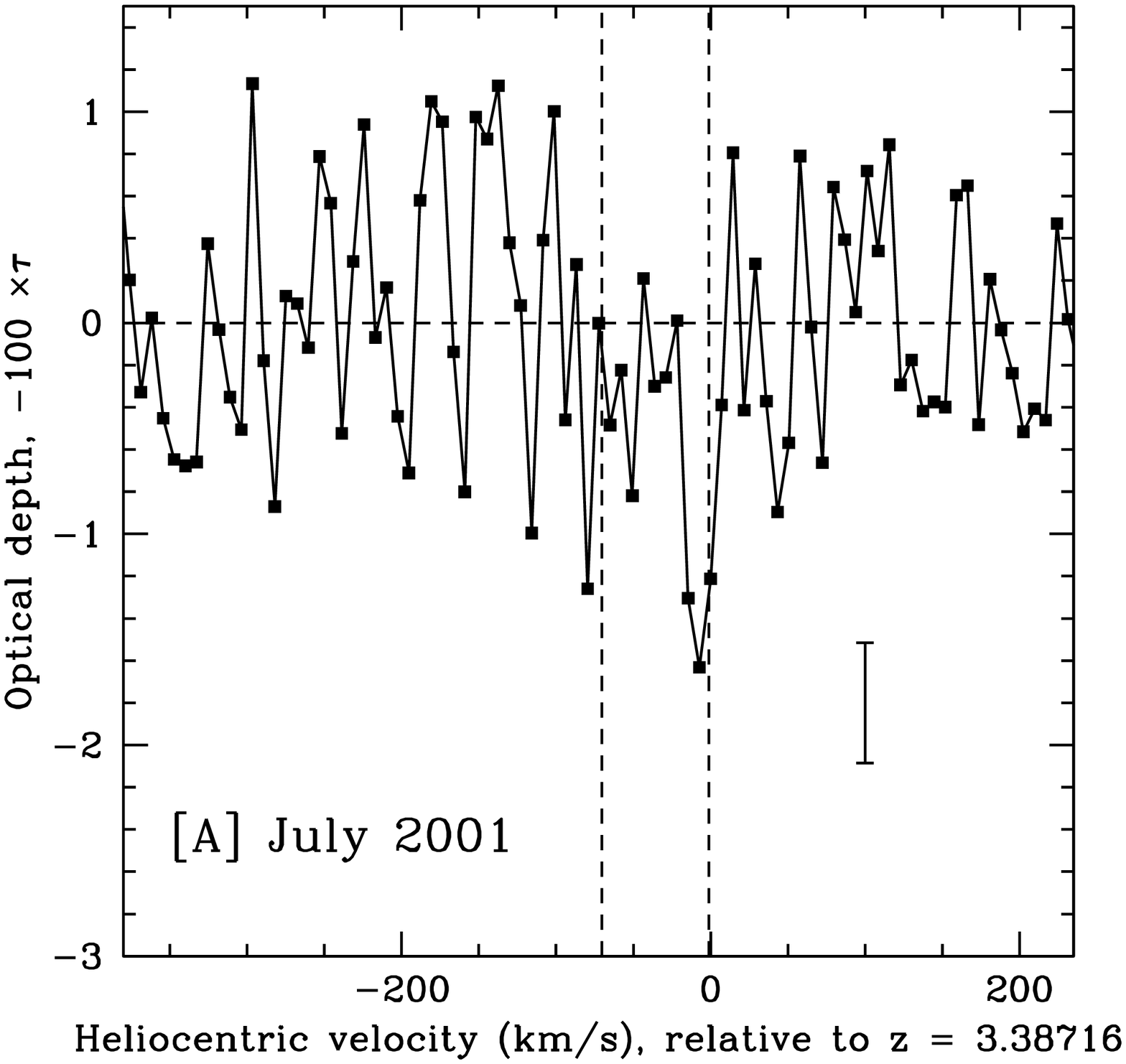,width=3.4in,height=3.4in}
\epsfig{file=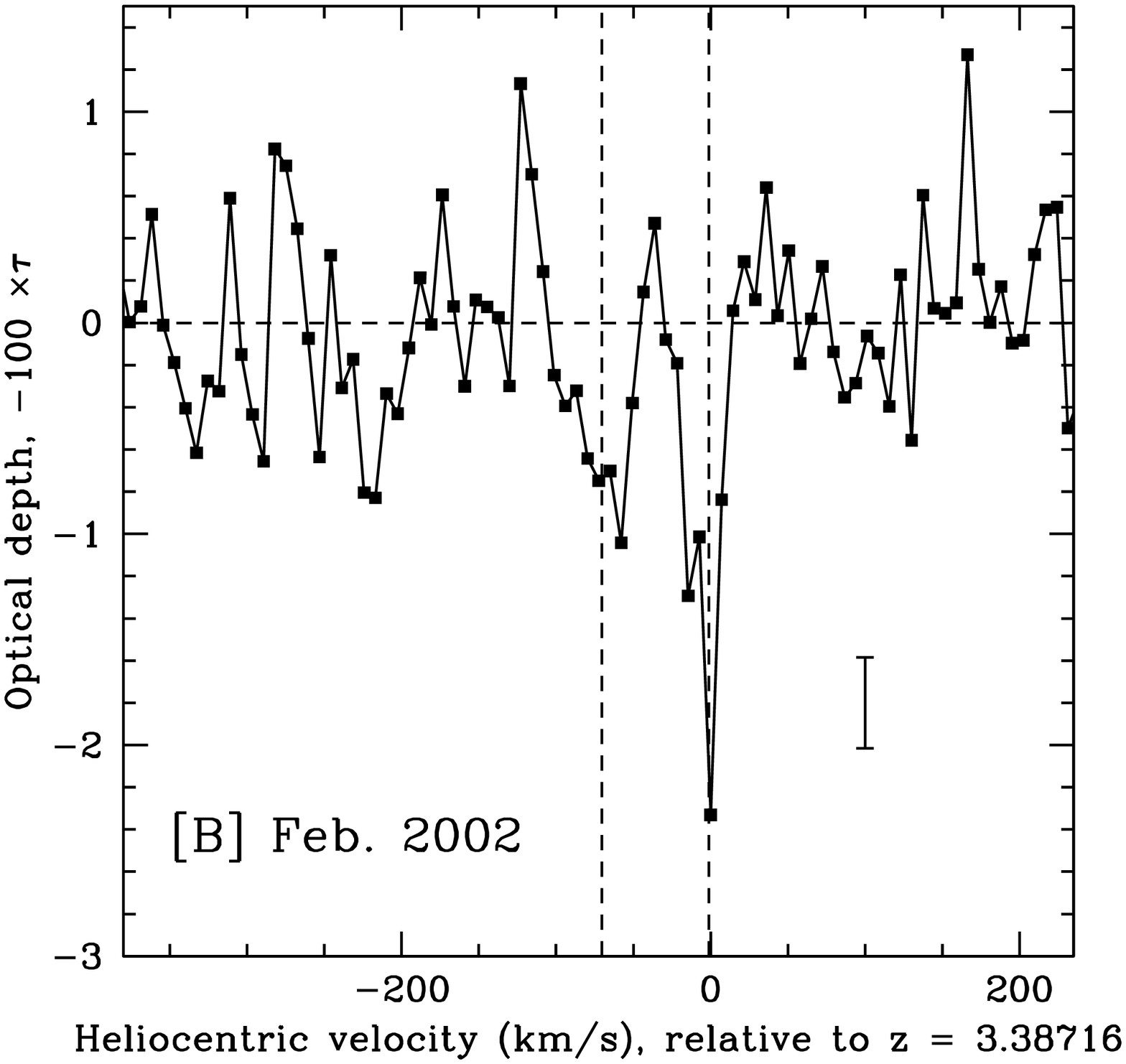,width=3.4in,height=3.4in}
\epsfig{file=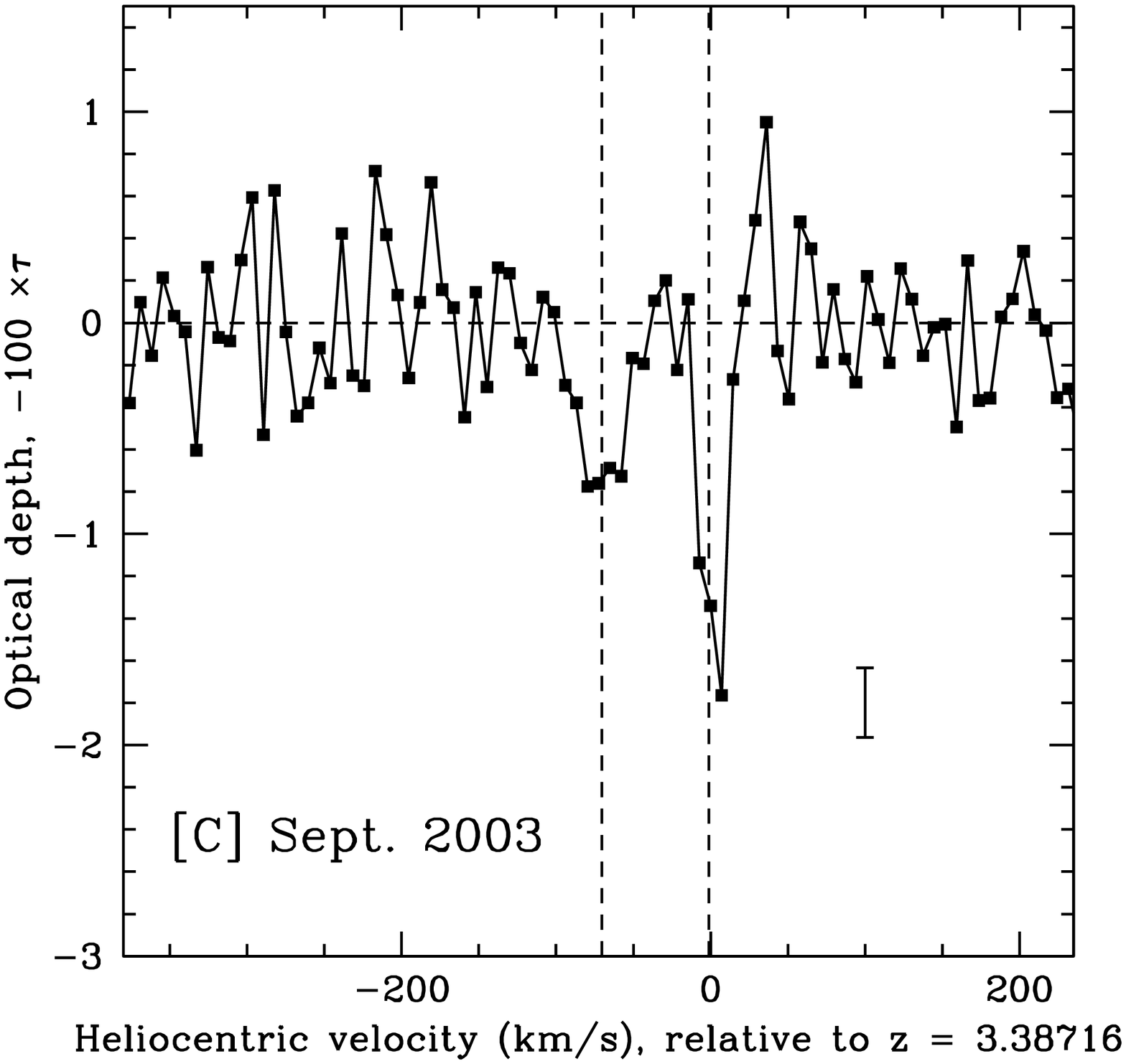,width=3.4in,height=3.4in}
\epsfig{file=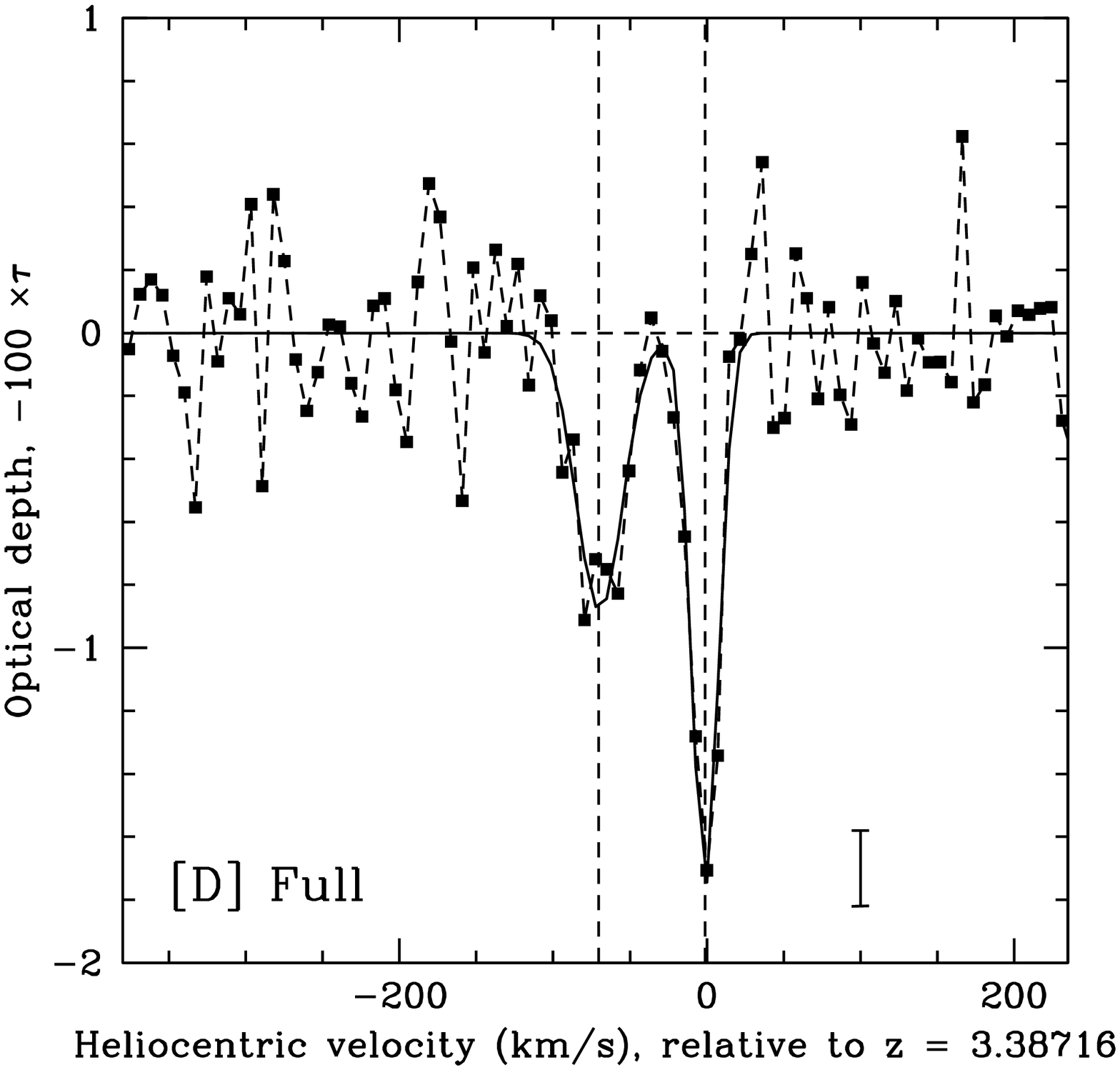,width=3.4in,height=3.4in}
\caption{GMRT HI 21cm spectra towards \pks~in [A]~2001~July, [B]~2002~February,
	and [C]~September~2003, at the original velocity resolution of $\sim 7.2$~\kms, 
        with optical depth ($-100 \times \tau$) 
	plotted against heliocentric velocity, in \kms, relative to 
	$z = 3.38716$. The main absorption component, close to zero
	velocity, is visible in all three spectra while the weaker component, 
        at $\sim -70$~\kms, can be seen in spectra [B] and [C], 
	which have higher sensitivity. Panel~[D] shows the final spectrum, after
        averaging the three spectra from [A]--[C]. The solid line in [D] shows 
        the 2-Gaussian fit to the spectrum.  The dashed vertical lines in each 
	panel indicate the peaks of the two Gaussians of Table~1, while the error bar 
 	on the right side indicates the $1 \sigma$ noise on each spectrum.}
\label{fig:fig1}
\end{figure*}

The $z \sim 3.39$ absorber towards \pks~\citep{white93} is one of the highest 
redshift DLAs known towards a radio-loud QSO and also has a very high HI 
column density  ($\NHI \sim 1.8 \times 10^{21}$~\cm; \citealt{ellison01}). 
Previous searches for 21cm absorption have yielded conflicting results, with two 
tentative detections (with substantially different redshifts, line depths and 
velocity spreads; \citealt{bruyn96,briggs97}) and two non-detections at similar 
sensitivity  \citep{briggs97, kanekar97}, all using different radio telescopes.  
High resolution Keck spectroscopy found no metal lines at either of the putative 
21cm redshifts, suggesting that the 21cm detections might have been spurious 
\citep{ellison01}. All the radio studies found evidence for a high spin temperature 
in the DLA, interpreted as a large fraction of warm HI. In contrast, the strong 
CII* lines in the Keck spectrum were used by \citet{wolfe03b} 
to argue that the DLA contains a large fraction of cold HI. We report here 
on new deep Giant Metrewave Radio Telescope (GMRT) observations of the DLA 
in the redshifted HI 21cm line.

\section{Observations and Data Analysis}
\label{sec:obs}

\begin{figure*}
\centering
\epsfig{file=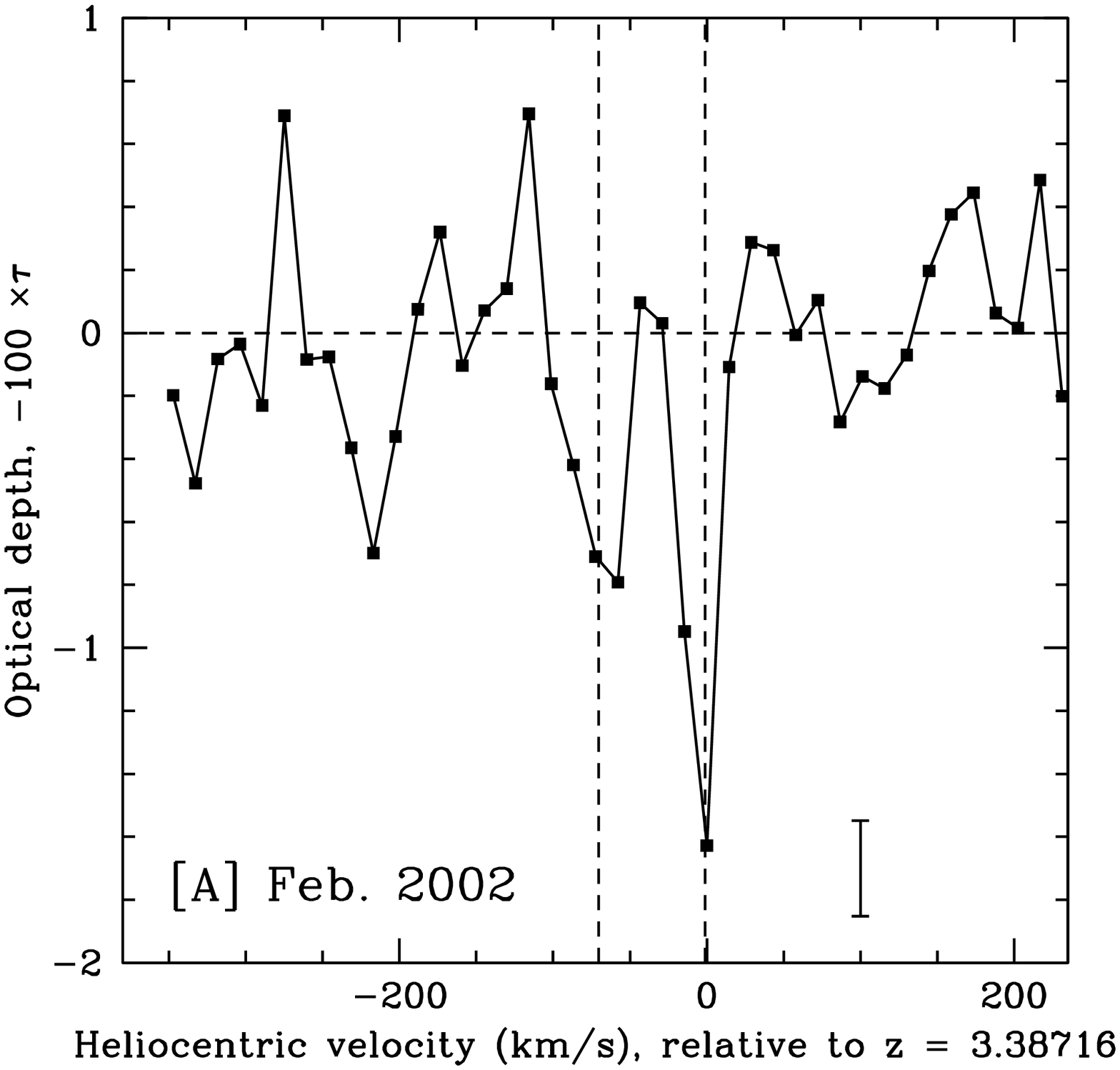,width=3.4in,height=3.4in}
\epsfig{file=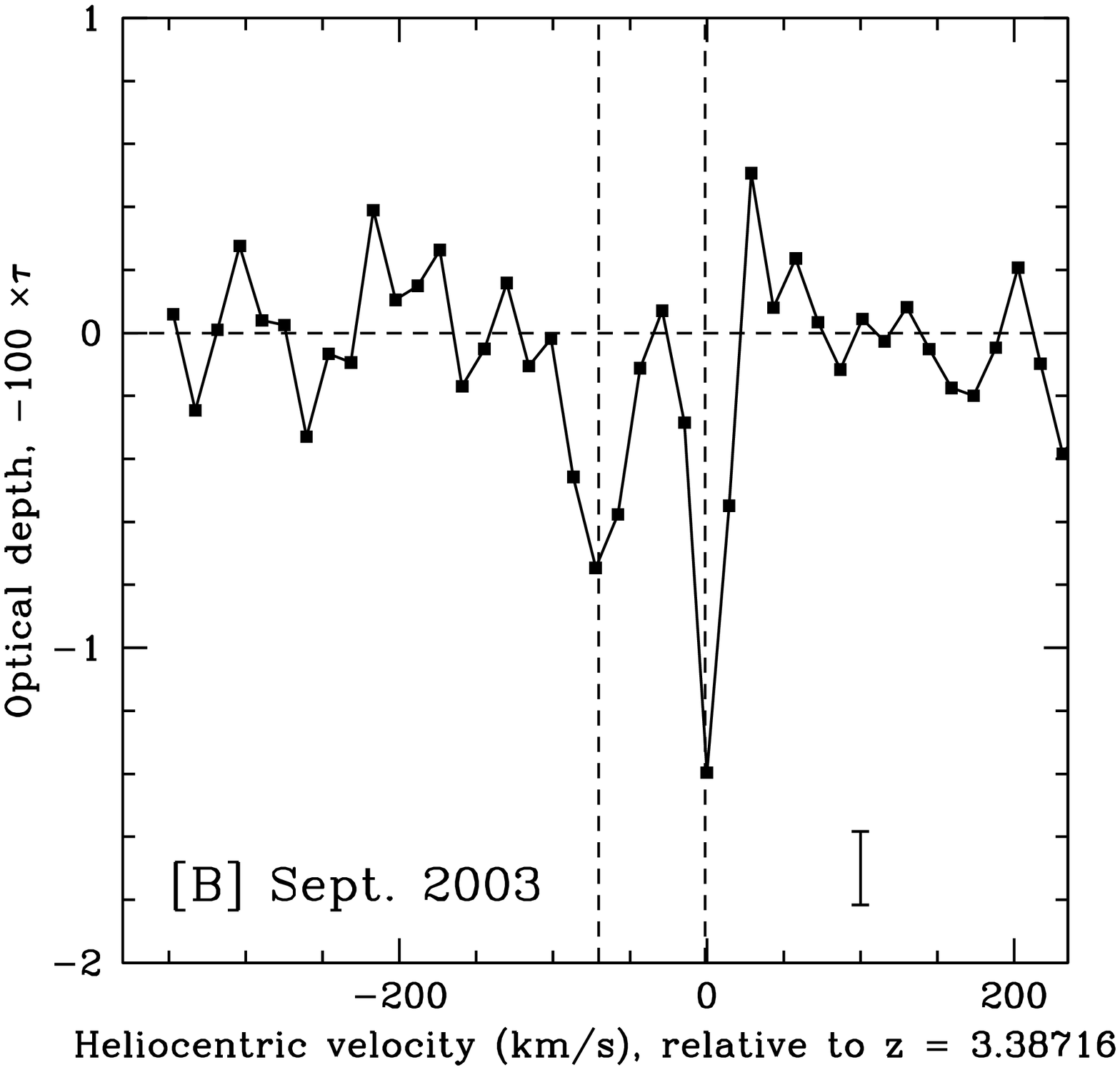,width=3.4in,height=3.4in}
\caption{GMRT Hanning-smoothed and re-sampled HI 21cm spectra towards \pks~from 
	[A]~2002~February, and [B]~September~2003, with optical depth ($-100 \times \tau$) 
	plotted against heliocentric velocity, in \kms, relative to 
	$z = 3.38716$. The spectra have a velocity resolution of $\sim 14.4$~\kms. 
        The weaker component, at $\sim -70$~\kms, is more clearly visible in both panels. 
        The dashed vertical lines in each panel indicate the peaks of the two 
	Gaussians of Table~1. The error bar on the right side of each panel indicates the 
	$1 \sigma$ noise on the spectrum.}
\label{fig:fig2}
\end{figure*}

\pks~was observed with the GMRT on multiple occasions, in 2001~January, 2001~July,
2002~February and 2003~September. The GMRT FX~correlator was used as the backend 
in all runs, with a 1~MHz bandwidth divided into 128 channels centred at the 
expected redshifted 21cm line frequency, giving a velocity resolution of $\sim 7.2$~\kms.
The first three runs used the upper sideband of the correlator while the session in 
2003 used the lower sideband. 19 antennas were available for the observations in 
2001 January, 24 in July 2001, 28 in 2002 February and 27 in 2003 September. The varying 
UV-coverage is not a concern as \pks~is unresolved by the longest GMRT baselines.

Observations of the standard calibrators 3C48 and 3C147 were used to 
calibrate the absolute flux density scale and the shape of the passband, while 
the compact source 0204+152 was used as the phase calibrator in all runs. 
The flux density of 0204+152 was measured to be within a few percent 
of 6~Jy, the 327-MHz value quoted in the Very Large Array (VLA) calibrator 
manual, in all cases except in 2001~July, when it was found to be $5.62 \pm 0.04$~Jy.
Note that the GMRT does not currently have online measurements of 
the system temperature; our experience indicates that the flux density scale
is reliable to $\sim 15$\% in this observing mode.


Data from the different runs were converted to FITS and analysed separately 
in classic AIPS, using standard procedures. After initial editing and gain and bandpass 
calibration, around 50 channels (chosen to be away from the expected line location) 
were averaged together into a ``channel-0'' dataset and then used to obtain a 
continuum image of the field out to a radius of $\sim 1$~degree, well 
beyond the half power of the GMRT primary beam. Data from different runs
were imaged independently to test for source variability. The image was
made by sub-dividing the field into 37~facets in all cases, to correct for 
the non-coplanarity of the array, and then used to self-calibrate the 
visibility data in an iterative manner until no further improvement in 
the image was obtained. This procedure typically converged after a few rounds 
of phase self-calibration, followed by two or three rounds of amplitude-and-phase 
self-calibration, with intermediate editing of corrupted data.

The task UVSUB was next used to subtract all continuum emission from the multi-channel 
U-V data and the residual visibilities inspected for any radio frequency interference 
(RFI) on all baselines,  using the task SPFLG to view the time-frequency plane. 
All detected RFI was edited out. The continuum-subtracted U-V data were shifted to 
the heliocentric frame (using the task CVEL) and imaged in all channels, and a 
spectrum then extracted from the image cube at the location of PKS~0201+113. Spectra 
were also extracted at the positions 
of other sources in the field to test for RFI. The data of 2001~January
were found to be affected by low-level RFI and will hence not be discussed further; data from 
other epochs were found to be clean. The total on-source times were $\sim 3$, $7$
and $10$~hours in 2001~July, 2002~February and 2003~September, respectively, 
after all editing. 

The 323~MHz flux density of PKS~0201+113 was measured to be 348~mJy, 417~mJy 
and 422~mJy in 2001~July, 2002~February and 2003~September, from the GMRT images. 
The typical RMS noise  in these images, away from bright sources, is $\sim 0.35 - 0.55$~mJy/Bm. 
While the flux density in 2001~July is $\sim 17$\% lower than in the 
other two runs, flux densities of other sources in the field (and the calibrator) 
were also lower by similar fractions; the GMRT data thus do not show any internal 
evidence for source variability. \citet{bruyn96} obtained a flux density of 
$(350 \pm 10)$~mJy with the Westerbork Synthesis Radio Telescope (WSRT) in 
1991~February, marginally consistent with the GMRT values (given the $\sim 15$\% 
uncertainty in the GMRT flux scale); we were unable to compare flux densities of 
other sources in the 0201+113 field for the WSRT dataset. \citet{briggs97} measured 
$(290 \pm 5)$~mJy in 1991~October with the VLA, significantly 
lower than the GMRT value. However, flux densities of other compact sources in the 
0201+113 field in the VLA image are also lower (typically, by $\sim 30\%$) than those 
in the GMRT image of 2003~September. Given the consistency of the GMRT and VLA measurements 
relative to field sources, we feel that the differences in measured flux density are unlikely 
to be caused by variability in 0201+113 itself. The GMRT flux density scale is likely to 
be correct, as the flux density of the calibrator, 0204+152, is in good agreement with that 
in the VLA calibrator manual. We will henceforth use a flux density of 422~mJy (from 2003~September) 
as the GMRT measurement.

%

We also obtained a 328~MHz image of \pks~with the Very Long Baseline Array (VLBA) 
in 2002, as part of a larger program to estimate the covering factors
of quasars with foreground DLAs. The observations and data analysis will be 
described elsewhere; we merely note here that these data were analysed in the 
AIPS and DIFMAP packages, using standard procedures. The total on-source time 
was $\sim 2$~hours.


\section{Spectra and Results}
\label{sec:spec}

The final spectra obtained in 2001~July, 2002~February and 2003~September
are shown in panels [A]--[C] of Fig.~\ref{fig:fig1}, with optical depth (in
units of $-100 \times \tau$) plotted as a function of heliocentric velocity,
relative to $z = 3.38716$. The spectra are at the original velocity resolution 
of $\sim 7.2$~\kms~ and have a root-mean-square (RMS) noise of 0.0057, 0.0043 
and 0.0033 per channel, respectively, in units of optical depth (indicated by 
the error bars on the right side of each panel). The main absorption component, 
at $\sim -1.0$~\kms, can be clearly seen in all the spectra while a second, weaker, 
component is visible at $\sim -70$~\kms~in the higher sensitivity spectra of panels 
[B] and [C]. The latter two spectra are shown in Figs.~\ref{fig:fig2}[A] and [B], after 
Hanning-smoothing and re-sampling to a velocity resolution of $\sim 14.4$~\kms, so 
that the weaker component can be seen more clearly. The doppler shift due to the 
Earth's motion between the three runs (e.g. $\sim 42$~\kms~or $\sim 45$~kHz between 
2002~February and 2003~September) is far larger than the spectral resolution; 
both features are hence likely to be real.

\setcounter{table}{0}
\begin{table}
\label{table:fit} 
\begin{centering}
\begin{tabular}{|c|c|c|c|c|}
\hline
 	&& && \\ 
	& $\tau_{\rm max}^\star$& $\nu^\dagger$ & $z$ 	       & FWHM$^\ddagger$ \\ 
 	& $\times 10^{-3}$ &   MHz          &  	       &  (\kms) \\ 
 	&& && \\ 
\hline
 	&& && \\ 

1	& $17.6 (2.0)$   & $323.7640 (12)$  & $3.387144 (17)$ & $18.6 \pm 2.6$  \\ 
2	& $8.8 (1.5)$	  & $323.8400 (34)$  & $3.386141 (45)$ & $35.1 \pm 7.4$ \\
 	&& && \\ 
\hline
\end{tabular}
\caption{Parameters of the 2-Gaussian fit to the final GMRT spectrum. Notes : 
$^\star$Peak optical depth, using the GMRT flux density of 422~mJy. 
$^\dagger$Heliocentric frequency. $^\ddagger$Full width at half maximum.}
\end{centering}
\end{table}

Fig.~\ref{fig:fig1}[D] shows the final GMRT spectrum towards PKS~0201+113, 
obtained from a weighted average of the three optical depth spectra in panels 
[A]--[C] of Fig.~\ref{fig:fig1}. The RMS noise on this spectrum is 0.0024 per 7.8~kHz 
channel (in optical depth units) while the equivalent width of the absorption 
is $\int \tau_{obs} {\rm dV} = (0.714 \pm 0.017)$~\kms.  The solid line 
in Fig.~\ref{fig:fig1}[D] shows a two-Gaussian fit to the spectrum; the 
parameters of the fit are summarised in Table~1.

\begin{figure}
\centering
\epsfig{file=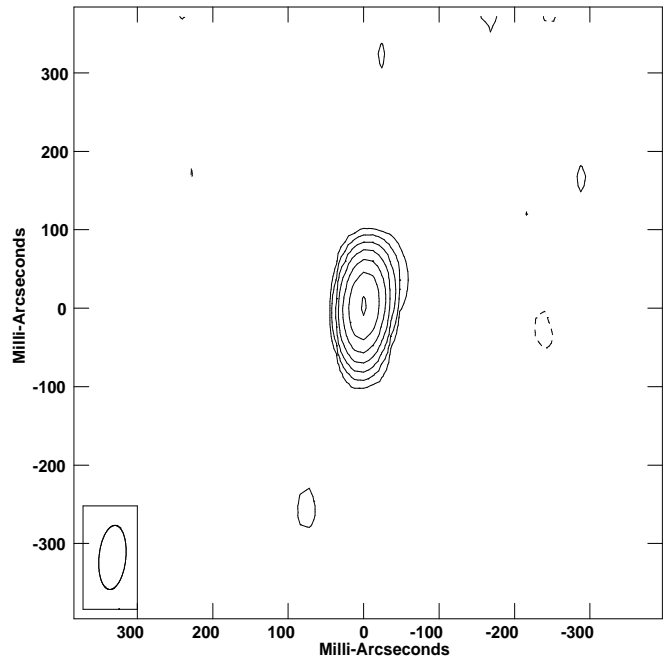,height=3.4in,width=3.4in}
\caption{VLBA 328~MHz continuum image of \pks, with a resolution of 
$78 \times 38$~mas. The contours are at $7.5 \times (-1,1,2,...,64)$~mJy.}
\label{fig:vlba}
\end{figure}

The 328~MHz VLBA image of \pks, shown in Fig.~\ref{fig:vlba}, has an angular 
resolution of $\sim 78 \times 38$~mas$^2$ (i.e. $\sim 588 \times 290$~pc$^2$ at 
$z \sim 3.39$, the DLA redshift). The source is well-fit by a single elliptical Gaussian 
component; we measure a core flux density of $\sim 292$~mJy and a deconvolved 
source size of $17.6 \times 6.6$~mas$^2$ (i.e. $\sim 132 \times 50$~pc$^2$ at $z \sim 3.39$\footnote{We 
use the standard $\Lambda$CDM cosmology throughout this paper, with $\Omega_m = 0.27$, 
$\Omega_\Lambda = 0.73$ and $H_0 = 71$~\kms~Mpc$^{-1}$}). 
A circular Gaussian model gives the same core flux density and an almost identical 
$\chi^2$ for the fit; the deconvolved source size is then $\sim 11.6 \times 
11.6$~mas$^2$ (i.e. $\sim 87.5 \times 87.5$~pc at the absorber redshift). 
We note that attempts to fit an additional component to the marginal north-west 
extension did not significantly alter the core flux density.

\section{Discussion}
\label{sec:discuss}

In this section, results of the present GMRT observations are initially compared with earlier 
21cm studies of the $z \sim 3.39$ DLA. The gas kinematics of the DLA are briefly discussed 
before we compare the 21cm and optical low-ionization metal profiles, as a probe of 
fundamental constant evolution. We next derive the spin temperature of the absorbing 
gas under the standard assumption that the average HI column density towards the radio 
core is the same as that measured towards the optical QSO from the damped Lyman-$\alpha$ line. 
The comparison between metal line and 21cm profiles finds evidence for small-scale structure 
in the absorbing gas across the radio core, suggesting that the average column density against
the radio core need not be the same as that towards the optical QSO. Finally, we discuss the 
HI temperature distribution in the light of the above results and two-phase models, and show 
that the gas must be predominantly warm in the absorber.

\subsection{A comparison with earlier 21cm absorption studies}
\label{sec:comparison}

\setcounter{table}{1}
\begin{table*}
\label{table:compare} 
\begin{centering}
\begin{tabular}{|c|c|c|c|c|c|c|c|c|c|c|}
\hline
 	&& && && &&&\\ 
Telescope & Epoch  &  Flux density & Resolution & RMS  & $\tau_{\rm max}$$^\dagger$ & Line depth$^\dagger$ & $z$ & FWHM   & References \\ 
          &        &   mJy & \kms     & mJy &  $\times 10^{-2}$ &      mJy             &     & (\kms) & \\ 
 	&& && && &&&\\ 
\hline
 	&& && && &&&\\ 
WSRT    & 1991  & $350 $ & $9.0$  & $7.0$ & $8.5 \pm 2.0$ & $30.0 \pm 7.0$         & $3.38699 (3)$   & $9 \pm 2$ & 1 \\ 
 	&& && && &&&\\ 
Arecibo & 1993  & $318$  & $4.5$  & $4.1$ & $3.7 \pm 0.8$ & $11.6 \pm 2.0$ & $3.38716 (7)$   & $23 \pm 5$ & 2\\ 
 	&& && && &&&\\ 
VLA	& 1991  & $290 $ & $11.1$ & $3.5$ & $1.5 \pm 1.2 $ & $4.2 \pm 3.4$ & -- & -- & 2 \\ 
 	&& && && &&&\\ 
ORT     & 1997  & $-$    & $5.6$  & $3.7$ & $1.3 \pm 0.6$ & $4.7 \pm 2.2$ & -- & -- & 3\\ 
 	&& && && &&&\\ 
GMRT    & 2001 -- 2003 & $422$  &$7.4$	& $1.0$ & $1.76 \pm 0.20$ & $7.43 \pm 0.84$ & $3.387144 (17)$ & $18.6 \pm 2.6$  & 4\\ 
        & 	       &        &       & & $0.88 \pm 0.15$ & $3.71 \pm 0.63$ & $3.386141 (45)$ & $35.1 \pm 7.4$  &  \\
 	&& && && &&&\\ 
\hline
\end{tabular}
\vskip 0.05in
\caption{A comparison between the 21cm spectra from different telescopes. $^\dagger$The quoted 
line and optical depths are from the Gaussian fits to the lines for the WSRT, Arecibo and 
GMRT spectra. For the VLA and ORT non-detections, the quoted line and optical depths are the values at 
323.764~MHz (ORT) and 323.770~MHz (VLA), after smoothing the spectra to resolutions of 
$\sim 16.8$~\kms~(ORT) and $\sim 22.2$~\kms~(VLA), respectively, and resampling. The ORT optical 
depth uses the WSRT flux density of 350~mJy, as the ORT observation did not have a reliable 
simultaneous flux density measurement. Note that the WSRT and VLA values have {\it not} been scaled 
to the GMRT flux density value of 422~mJy but are on the original flux scales.
References : (1)~\citet{bruyn96}; (2)~\citet{briggs97}; (3)~\citet{kanekar97}; (4)~this paper.}
\end{centering}
\end{table*}

\begin{figure}
\centering
\epsfig{file=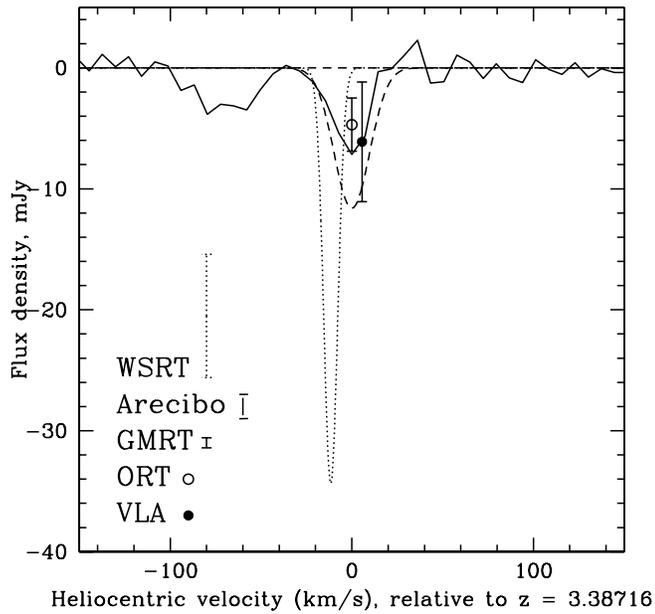,height=3.4in,width=3.4in}
\caption{A comparison between the final GMRT spectrum (solid line) and earlier Arecibo, WSRT, VLA 
and ORT results, with flux density (in mJy) plotted against heliocentric velocity, in km/s, 
relative to $z = 3.38716$. The dashed and dotted lines show the Gaussian fits to the earlier 
tentative Arecibo and WSRT detections of \citet{briggs97} and \citet{bruyn96}, respectively. The solid 
and open circles (and $1\sigma$ errors) indicate the values at 323.764~MHz (ORT) and 
323.770~MHz (VLA) \citep{briggs97,kanekar97} respectively, after smoothing to resolutions 
of $\sim 16.8$~\kms~(ORT) and $\sim 22.2$~\kms~(VLA). The VLA and WSRT line depths have been scaled 
to the GMRT flux density of 422~mJy for the comparison. The $1\sigma$ errors on the Arecibo, 
WSRT and GMRT fits are shown by the error bars to the right of each telescope name. 
See text for discussion.}
\label{fig:fig4}
\end{figure}

As noted in the introduction, the $z \sim 3.39$ DLA has been the focus of 
a number of 21cm absorption studies in the past, with very different results
\citep{bruyn96,briggs97,kanekar97}. Details of the different observations are 
listed in Table~2, with the line depths quoted in both optical depth and flux density,
while Fig.~\ref{fig:fig4} shows a visual comparison between the GMRT spectrum and 
the earlier results. Note that the comparison in the figure is in flux density units 
(rather than in optical depth), as the continuum flux density estimates at Arecibo and the ORT 
are unreliable due to the very large beamwidths and the associated confusion at the low observing 
frequency. The VLA and WSRT line depths have been scaled to the GMRT flux density of 
422~mJy for the comparison.

It is clear from Fig.~\ref{fig:fig4} and Table~2 that the VLA and ORT non-detections
are consistent with the GMRT detection within the $1\sigma$ errors; note that both spectra 
showed weak absorption (within the noise) close to the frequency of the main GMRT component 
($\sim 323.76$~MHz). Next, the fit to the Arecibo feature of \citet{briggs97} is in excellent 
agreement with the stronger GMRT 21cm component in both redshift and FWHM. However, the peak 
optical depth is lower in the GMRT spectrum, discrepant at $\sim 2.4 \sigma$ level. Similarly, 
the Arecibo peak optical depth differs at $\sim 2 \sigma$ level from the ORT limit 
\citep{kanekar97} and is only in marginal agreement with the VLA non-detection \citep{briggs97}. 
A plausible cause of the optical depth discrepancy is an incorrect continuum flux density in 
the Arecibo data, a not-unusual problem with single-dish measurements. However, the line depth, 
in flux density units, in the Arecibo spectrum is also discrepant from the GMRT measurement at 
$\sim 2 \sigma$ level, implying that the difference in optical depths is not solely due to an 
error in the continuum flux density. As noted by \citet{briggs97}, low-level RFI was present 
throughout the Arecibo spectrum (again, not uncommon for single-dish observations), resulting 
in non-Gaussian behaviour and possible systematic errors. This could well result in an 
incorrect determination of the Arecibo line depth and, hence, an over-estimate in the optical depth. 

The only significantly discrepant result in Fig.~\ref{fig:fig4} and Table~2 is the
WSRT feature of \citet{bruyn96}, which is inconsistent with all the other data. The 
WSRT feature is unresolved and was detected 
at only $\sim 4.5 \sigma$ level in one polarization \citep{bruyn96}; further these data 
have the lowest sensitivity of all the measurements. Given the difficulty of these 
low frequency observations in the presence of RFI, the relatively low statistical 
significance and the disagreement with all the other observations, the WSRT feature 
appears likely to be an artefact.

One possibility is that the above differences between the WSRT, Arecibo and GMRT 
features might be due to real changes in the optical depth of the 21cm absorption, 
as has been seen in two low~$z$ DLAs, towards AO~0235+164 \citep{wolfe82} and 
PKS~1127--145 \citep{kanekar01c}. However, the GMRT spectra show no internal evidence 
for optical depth variability between the different observing epochs (from 2001 to 2003) 
and are also consistent with the VLA and ORT spectra (of 1991 and 1997), making optical 
depth variability an unlikely possibility. We hence conclude that (1) the WSRT feature 
of \citet{bruyn96} is likely to be an artefact, (2) the GMRT, ORT, VLA and Arecibo spectra 
are consistent, if one assumes an incorrect estimate of the line depth in the Arecibo 
spectrum (note that the Arecibo spectrum is only discrepant at $\sim 2 \sigma$ level 
even without this assumption, with the FWHM and redshift in good agreement), and (3) there is 
no strong evidence for variability in the 21cm optical depth in the $z \sim 3.39$ DLA.
Of course, it is obvious from the RMS values of Table~2 that the second, weaker, 
feature seen in the GMRT spectrum is far too weak to have been detected in any of the earlier 
observations.

\subsection{Gas kinematics}
\label{sec:kinematics}

The 21cm absorption profile of Fig.~\ref{fig:fig1}[D] has a total velocity spread 
(between nulls) of $\sim 115$~\kms, the largest 21cm velocity width seen in 
any $z \gtrsim 2$ DLA. In fact, this width is comparable to values seen in low~$z$ DLAs 
identified as bright spiral galaxies (for comparison, see Table~3 in \citealt{kanekar03}).
The only low~$z$ DLA with a 21cm velocity width larger than $\sim 100$~\kms~and {\it not} 
associated with a spiral disk galaxy is the $z \sim 0.3127$ absorber towards PKS~1127--145 
\citep{lane98,kanekar01c}. Interestingly enough, \citet{ellison01} find the gas kinematics 
of  the $z \sim 3.39$ DLA towards \pks~to be chaotic, with no evidence for an edge-leading 
asymmetry in the optical lines (a signature of a rotating disk; \citealt{prochaska97}). 
Instead, the low-ionization metal lines show a large number of spectral components over a velocity 
spread of $\sim 270$~\kms. Similarly, both 21cm components in the $z \sim 3.39$ DLA are 
quite wide (see FWHMs in Table~1), perhaps due to multiple blended sub-components. 

\subsection{Fundamental constant evolution}
\label{sec:alpha}

Comparisons between the redshifts of different spectral transitions can be used to 
probe evolution in the values of the fundamental constants such as the fine structure 
constant $\alpha$, the electron-proton mass ratio $\mu \equiv m_e/m_p$, the
proton gyromagnetic ratio $g_p$, etc. A variety of approaches employing different 
transitions have been used, constraining changes in the above constants to be 
$\lesssim 10^{-5}$ (e.g. \citealt{murphy03,anand04,kanekar05,reinhold06}). 
One such technique uses optical metal lines and the 21cm line to probe changes 
in the quantity $X \equiv g_p \mu \alpha^2$ \citep{wolfe76}. However, the velocity 
structure can be very different in the optical and radio lines, making it difficult 
to identify a single absorption redshift for the comparison, especially in cases
where the profiles are complex, with multiple spectral components. While one might 
compare the redshifts of the strongest absorption component in the 21cm and metal 
lines \citep{tzanavaris06}, there is no physical reason why the strongest optical 
and 21cm absorption should arise in the same spectral component; this depends on 
local physical conditions in the absorbing clouds. The ``strongest-component'' 
approach thus introduces an extra source of systematic error that could lead to 
a bias in favour of a detection in small absorption samples. Conversely, while a 
comparison between the redshifts of ``nearest-neighbour'' absorption components 
in the 21cm and optical lines \citep{kanekar06} might be viewed as a test of the 
null hypothesis, it has the opposite bias, i.e. towards a non-detection of evolution. 

\begin{figure*}
\centering
\epsfig{file=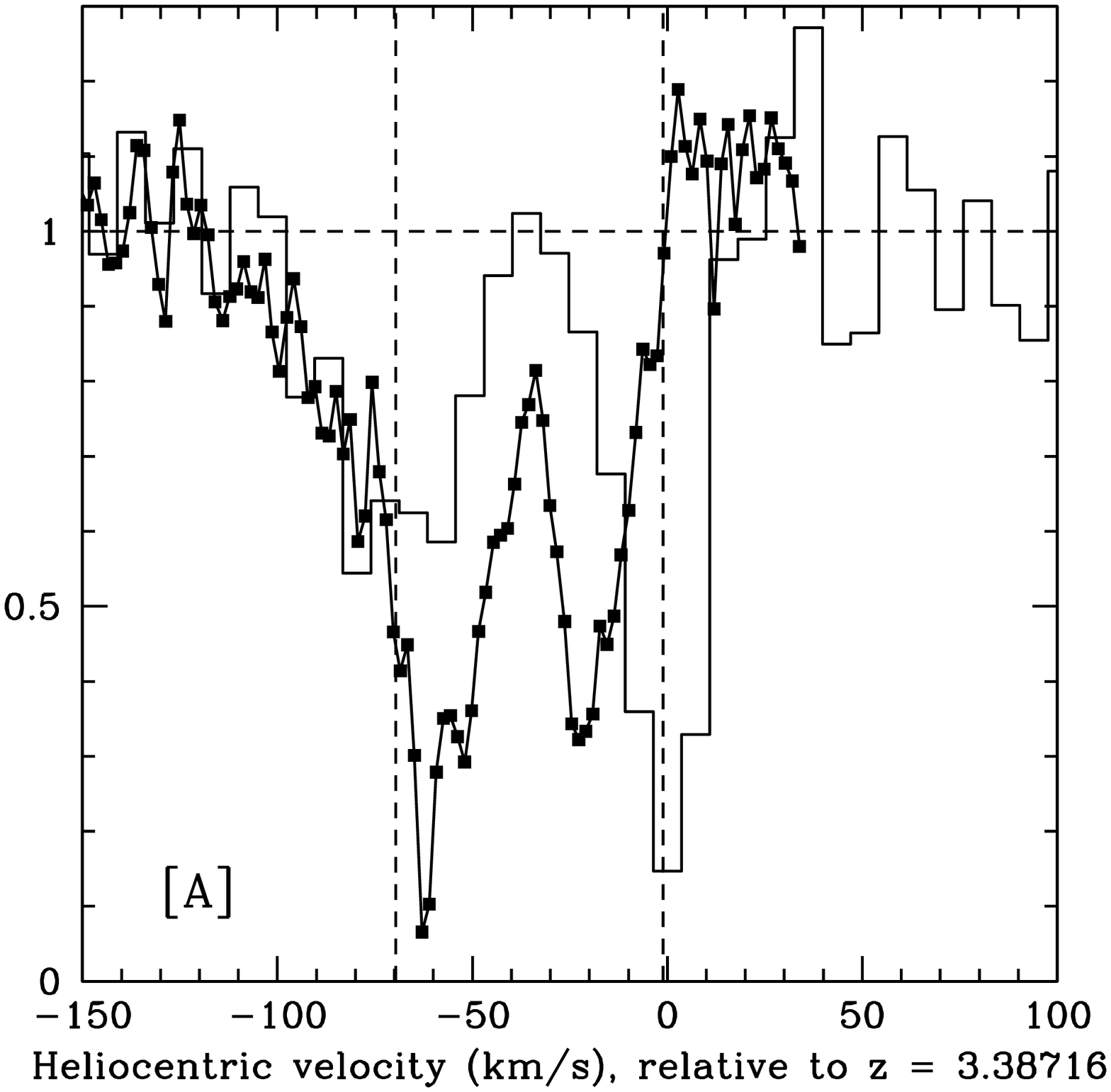,height=3.3truein,width=3.3truein}
\epsfig{file=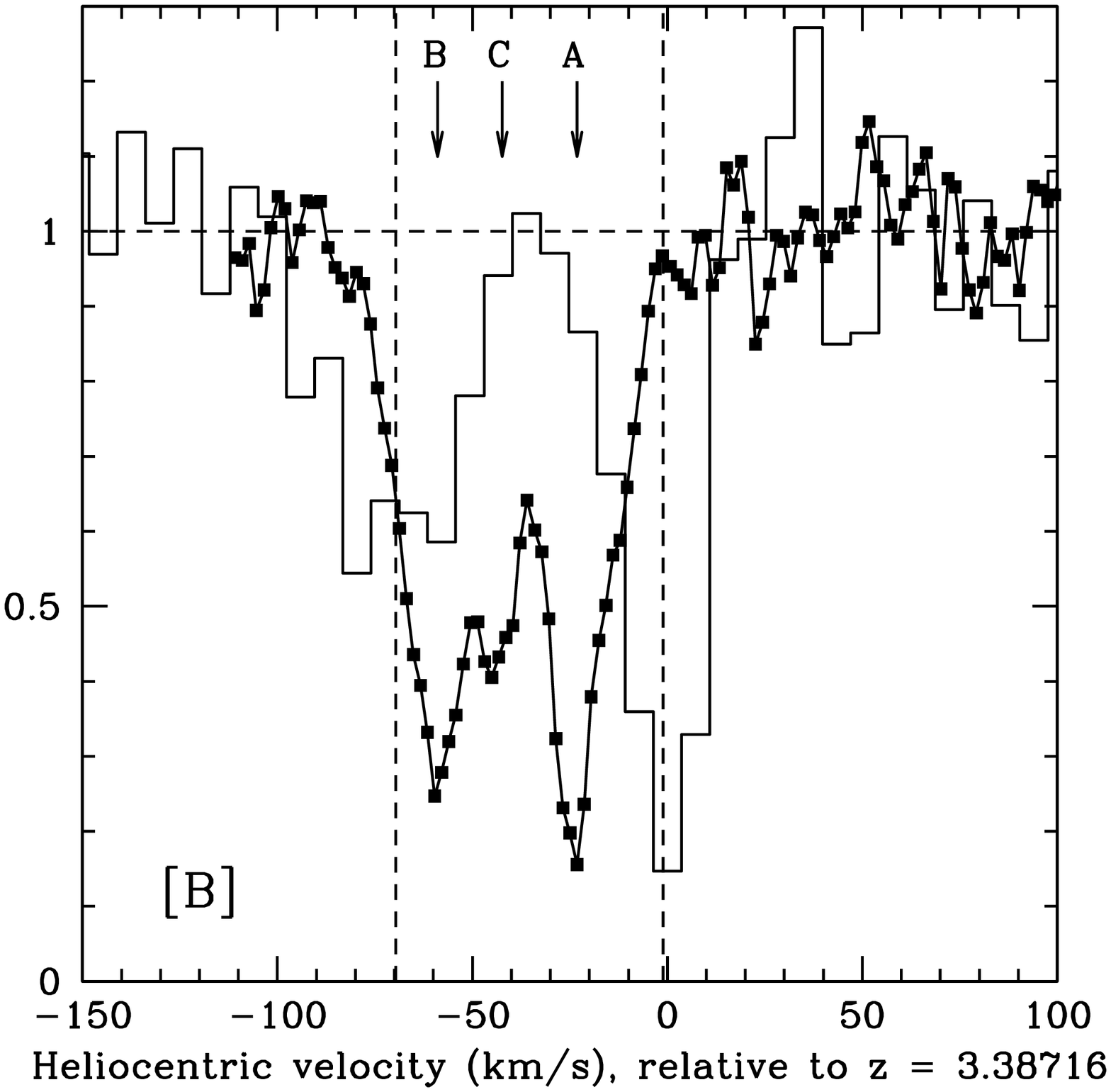,height=3.3truein,width=3.3truein}
\caption{A comparison between the GMRT HI~21cm profile (histogram) and
the [A]~FeII~$\lambda$1122 and [B]~CII*~$\lambda$1335 Keck-Hires profiles 
(solid squares). The  spectra are in arbitrary units and plotted against
heliocentric velocity, in \kms, relative to $z = 3.38716$. The dashed 
vertical lines in each panel indicate the redshifts of the two 21cm 
components from the 2-Gaussian fit. The three vertical arrows in [B] indicate 
the redshifts of the three components (A, B and C) obtained from the fit to 
the CII* line.  See text for discussion.}
\label{fig:hi-metal}
\end{figure*}

The $z \sim 3.39$ DLA towards \pks~is the highest redshift absorber in which 21cm and 
optical lines have both been detected and is thus an obvious candidate for studies
of changes in $X \equiv g_p \mu \alpha^2$. Unfortunately, both the 21cm and optical 
profiles are complex, extending over $\gtrsim 100$~\kms, with multiple components. 
Fig.~\ref{fig:hi-metal} shows the GMRT 21cm profile (histogram) overlaid on the 
unsaturated FeII-$\lambda$1122  and CII*-$\lambda$1335 lines detected in the 
Keck-HIRES  spectrum of \citet{ellison01}. It can immediately be seen that even 
the strongest optical component is not uniquely determined; the absorption at 
$\sim -65$~\kms~is stronger in the FeII-$\lambda$1122 transition while that at 
$\sim -25$~\kms~is stronger in the CII* line. We will use the CII* transition for the 
comparison, as this is expected to primarily arise in cold gas  (\citealt{wolfe03b}; 
hereafter W03), which also gives rise to the 21cm absorption. However, similar results 
are obtained on using other low-ionization metal species.

The CII* absorption has three components, at $z_A = 3.386821 (7)$,  $z_B = 3.386297 (10)$ 
and $z_C = 3.386541 (10)$, in order of decreasing strength (the redshifts are indicated 
by the arrows in Fig.~\ref{fig:hi-metal}[B]),  while the 21cm profile has two components, 
the main one at $z_1 = 3.387144 (17)$ and the secondary at $z_2 = 3.386141 (45)$. A 
strongest-component comparison between the  21cm and CII* lines (i.e. between $z_A$ and 
$z_1$) gives 
$\lsb \Delta X/X \rsb = \lsb \Delta z/(1 + \bar{z})  \rsb = (-7.4 \pm 0.4) \times 10^{-5}$, 
where $\Delta z = z_{\rm CII*} - z_{\rm 21}$ and $\bar{z} = (z_{\rm 21} + z_{\rm CII*})/2$.  
If this value of $\lsb \Delta X/X \rsb$ is due to changes in $\dal$ or $\dmu$, 
it would require $\dal$ or $\dmu$ to be nearly an order of magnitude larger than 
values obtained from other methods probing a similar redshift range\footnote{Note 
that  changes in $g_p$ are expected to be much smaller than those in $\alpha$ 
and $\mu$ \citep{calmet02,flambaum04}} \citep{murphy03,anand04,reinhold06}. 
Further, \citet{tzanavaris06} obtained $\lsb \Delta X/X \rsb = (0.63 \pm 0.99) \times 10^{-5}$ 
over $0.23 < z < 2.35$ from a strongest-component comparison between 21cm and 
optical lines in nine absorbers, again significantly smaller than the present value.
This suggests that the difference between the redshifts of the strongest components 
towards \pks~ stems from intrinsic differences between the radio and optical sightlines, 
discussed further in the next section.  

Alternatively, one might compare the redshifts of the weaker 21cm feature and 
its nearest-neighbour CII* component (i.e. component~B); formally, this yields 
$\lsb \Delta X/X \rsb = (+3.6 \pm 1.0) \times 10^{-5}$. However, these CII* and 
21cm components have very different velocity widths (FWHMs of $\sim 13$ and 
$\sim 21$~\kms, respectively), suggesting that at least part of the absorption 
does not arise in the same gas. A nearest-neighbour comparison between these 
optical and 21cm components is thus unlikely to yield reliable results. We 
hence conclude that, despite its high redshift, the $z \sim 3.39$ DLA is unlikely 
to be useful for the purpose of probing fundamental constant evolution.

\subsection{Physical conditions in the absorber}
\label{sec:tspin}

The HI 21cm absorption in DLAs has generally been regarded as arising in a two-phase medium, 
with both cold (CNM) and warm (WNM) gas, similar to that seen in the Galaxy.
Observational support for such a model has been found in  the case of two low $z$ DLAs 
\citep{lane00,kanekar01b}. In the context of such two-phase models, the fact that 
high $z$ DLAs generally have high spin temperatures (e.g. \citealt{bruyn96,carilli96,
kanekar01a,kanekar03}) implies that the bulk of the gas is in the warm phase. Recently,
however, W03 used the observed CII* absorption in DLAs as a tracer of the cooling rate 
to argue that DLAs with strong CII* lines must have a substantial fraction of their gas 
in the CNM phase [see, however, \citet{anand05}]. W03 ruled out models in which all 
of the observed CII* absorption arises in the WNM, finding that strong CII* absorption
requires that the line-of-sight includes some CNM [see also  \citet{wolfe04}]. They 
further argued that the multiple lines of evidence suggesting that HI in high~$z$ DLAs
consists  primarily of warm gas (e.g. \citealt{liszt02,norman97,kanekar03})
are inconclusive, and  suggested instead  that a two-phase model with a CNM fraction 
of $50$\% towards the optical QSO is consistent with the observed bolometric 
luminosity in DLAs.

    As noted by W03, the $z \sim 3.39$ absorber towards \pks~is one of only two high $z$ DLAs 
that has both strong CII* absorption (suggesting a high CNM fraction in the W03 model), 
as well as a high 21cm spin temperature. W03 resolved this apparent contradiction by 
arguing that the background radio core must subtend a linear diameter of $\gtrsim 40$~pc 
at the DLA, while the CNM ``clouds'' in the absorber have linear sizes of $\lesssim 10$~pc. 
They suggested that the strong CII* absorption could be consistent with a lack of 21cm 
absorption if the CII* absorption comes from a number of small CNM clouds lined up 
against the AU-sized optical QSO. We revisit this issue in the light of the new 
observations presented in the preceding sections. 

A convenient starting point for this discussion is the general relation between 21cm optical depth
$\tau_{21}$, total HI column density $\NHI$ (in \cm) along the path, and spin temperature $\ts$ (in K)
in the optically thin limit \citep{rohlfs04},

\begin{equation}
\label{eqn:tspin1}
\NHI = 1.823 \times 10^{18} \ts \int \tau_{21} {\rm dV} \;\; ,
\end{equation}
where the integral is over velocity, in \kms. The spin temperature thus obtained is the 
column-density-weighted harmonic mean of the $\ts$ values of different HI phases along 
the path. As such, it does not give the kinetic temperature of any individual phase but 
instead contains information on the distribution of gas in different temperature phases 
along the line of sight.

Next, the total flux density of the source measured by a short-baseline interferometer such as 
the GMRT or the VLA may be much larger than the flux density that is covered by the foreground 
absorbing ``cloud'' (e.g. \citealt{briggs83}). This is especially true at the high redshift of 
the present observations; for example, the GMRT synthesized beam in the dataset of September~2003 
is $\sim 12.2'' \times 10.0''$, subtending spatial dimensions of $\sim 100 \times 75$~kpc at 
$z \sim 3.39$, far larger than the size of a typical galaxy. In such cases, the {\it measured} 
21cm optical depth $\tau_{obs}$ will be smaller than the ``true'' 21cm optical depth, $\tau_{21}$. 
If only a fraction $f$ of the radio continuum is covered by the foreground absorbing gas, the 
``true'' 21cm optical depth is given by $\tau_{21} = (1/f)*\tau_{obs}$. For a damped absorber, 
where the HI column density is measured from the Lyman-$\alpha$ line towards the optical QSO, 
we then have

\begin{equation}
\label{eqn:tspin}
\NHI = 1.823 \times 10^{18} \lb \ts /f \rb \int \tau_{obs} {\rm dV} \;\; ;
\end{equation}
\noi $f$ is referred to as the 21cm covering factor of the DLA.

In addition to the assumption that the 21cm absorption is optically thin, 
Eqn.~\ref{eqn:tspin} contains two critical unknowns, the DLA covering factor 
$f$, and whether
the HI properties along the line of sight to the optical QSO can be applied 
to the much larger line of sight towards the more-extended radio quasar core.
The covering factor has been the subject of much discussion in the literature 
(e.g. \citealt{briggs83,curran05}). In principle, the ``true'' 21cm optical depth 
can be directly determined by using VLBI techniques to map the 21cm absorption 
(e.g. \citealt{lane00}); however, this is usually impossible due to the fairly poor 
frequency coverage and sensitivity of VLBI receivers. The alternative approach is to use 
VLBI observations at nearby frequencies to measure the fraction of flux density arising 
from the compact radio core. Even such 
measurements are rarely carried out in practice due to the technical difficulties in
such VLBI studies at low frequencies (e.g. \citealt{briggs83b}).
On the other hand, it is difficult, even in principle,
to determine whether the HI distribution is uniform across the extended background 
radio source [see, for example, the discussions in \citet{bruyn96} and W03]. 
Given that these two assumptions are generally untested in high $z$ DLAs, the 
case of PKS0201+113 is doubly unusual in that we have meaningful VLBI constraints on 
the size of the radio core, and, as discussed below, also find observational 
evidence for small-scale structure in the HI~21cm absorbing gas. 

For ease of comparison with earlier work, however, we first estimate the spin 
temperature of the $z \sim 3.39$ DLA with the usual assumption that the HI gas is
uniformly distributed across the radio core. If the 21cm absorber is a uniform 
slab covering the entire VLBI core, the covering factor $f$ is 
$S_{VLBA}/S_{GMRT} \sim (292/422) \sim 0.69$.
Further, $\NHI = (1.8 \pm 0.3) \times 10^{21}$~\cm~\citep{ellison01}, and
$\int \tau_{obs} {\rm dV} = (0.714 \pm 0.017)$~\kms, which gives $\ts \sim [(955 \pm 160) 
\times (f/0.69)]$~K. This is in the ``high'' range of $\ts$ values in DLAs 
\citep{kanekar03}. We note that the above $\ts$ value is significantly lower than 
the estimate $\ts > 3290$~K in \citet{kanekar03}. This is because the latter 
assumed $f \sim 1$ and a thermal velocity spread  (FWHM~$= 20$~\kms) which is
much smaller than the true velocity spread of $\sim 115$~\kms.

 We next consider the assumption that the gas is uniformly distributed across the
 radio core. If the radio source were indeed very compact ($\lesssim 10$~pc), strong 
interstellar scintillation would be expected at the low GMRT observing frequencies, 
giving rise to a variable source flux density. However, the GMRT observations found no 
evidence for flux density variability over time-scales of a few years.  It thus appears likely 
that the radio core is at least a few tens of parsecs in size at the redshifted 21cm frequency, far 
larger than the size of the optical core. Our VLBA observations cannot rule out this possibility 
as they only require the radio core to be smaller than $\sim 100$~pc, the size of the fitted 
circular or elliptical Gaussian models. 

The large difference in the sizes of the optical and radio cores implies that a comparison 
between the profiles of low ionization metal lines and the 21cm line would reveal the 
presence of small-scale velocity structure in the absorber over scales of tens of parsecs. 
This is because the low ionization metal lines trace the kinematics of the gas along the 
line of sight to the AU-sized optical QSO, while the 21cm line traces the kinematics 
along the much-larger ($\gtrsim 40$~pc) region lying in front of the radio core. 
No metal-line absorption can be seen in Figs.~\ref{fig:hi-metal}[A] and [B] 
at $z \sim 3.387144$, the redshift of the stronger 21cm component. As noted in 
Section~\ref{sec:alpha}, it is not possible to account for the large observed redshift 
difference between the 21cm and optical lines by fundamental constant evolution.  
Further, as pointed out by \citet{ellison01}, the 21cm peak redshift is at the edge of 
the absorption seen in even the saturated OI-$\lambda$1302 and CII-$\lambda$1334 transitions 
(see Fig.~5 and the discussion in \citealt{ellison01}). This suggests that the stronger 
21cm component does not arise along the line-of-sight to the optical core.  While it is 
possible that the component arises against the extended radio emission that is resolved out in 
the VLBA image, this would require the absorbing clouds to be either very smooth and 
extended or, if compact, to have an extremely high 21cm optical depth. The more likely 
situation is that one or more compact CNM clouds cover part of the radio core,
(but not the optical core) and give rise to this 21cm component. In contrast, 
the weaker 21cm component lies close to the deepest FeII-$\lambda$1122 and CII* absorption 
and fairly strong metal-line absorption is seen over most of its velocity spread. This is 
consistent with at least part of the latter 21cm component arising from gas that does lie in 
front of the optical emission. The absorbing gas thus indeed appears to have structure 
on the scale of the radio core. 

Despite the presence of this small-scale structure, it can be shown that if the 
HI is indeed in a two-phase medium as in the Galaxy, the WNM phase is likely to be 
dominant for the absorber as a whole, on the scales probed by the radio observations.
This is because the HI in such two-phase models consists of randomly-distributed dense 
CNM ``clouds'' surrounded by a diffuse, widespread WNM envelope. If we assume that the 
optical QSO lies along a line of sight for which the CNM fraction is $\sim 50$\% (i.e. 
the scenario favoured by W03), the remaining half of the HI column density along this 
sightline is WNM, i.e. $N_{\rm WNM} \sim 9 \times 10^{20}$~cm$^{-2}$.  The number density 
in stable WNM is $\lesssim 0.1$~cm$^{-3}$ \citep{wolfire95}; a WNM cloud with 
N$_{\rm HI} \sim 9 \times 10^{20}$~cm$^{-2}$ must thus have a linear  
extent of $\sim few$~kpc along the line-of-sight, far larger than the VLBA upper limit 
on the size of the radio core. The entire radio core must hence 
be covered by the WNM, with an average WNM column density of $\sim 9 \times 10^{20}$~\cm.
Given the negligible 21cm  optical depth of the WNM, the detected 21cm absorption components
of Fig.~\ref{fig:fig1}[D] must arise in the CNM. Using a nominal CNM spin temperature 
of $\sim 100$~K, the average CNM column density over the radio core (i.e. the core 
detected in the VLBA image) is $N_{\rm CNM} = 1.823\times10^{18}\times 100 \times (1/0.69) 
\times 0.714$, i.e.  $N_{\rm CNM} \sim 1.9\times10^{20}$~\cm. The average CNM fraction 
across the radio core is hence $N_{\rm CNM}/(N_{\rm CNM}+N_{\rm WNM}) \sim 0.17$. Thus, 
even if the line-of-sight to the optical QSO has a fortuitously high CNM fraction, the 
large-scale HI distribution must be dominated by the WNM.

We note that it is still possible to have a large fraction of the total HI in the CNM 
phase if the CNM were to be distributed in small ``clouds'' with velocity separations 
larger than the individual velocity widths, so that the absorption from each cloud lies 
below the GMRT 21cm detection threshold. As always for absorption studies, such a model, in 
which most of the CNM is ``hidden'', cannot be ruled out from the absorption data alone.

\section{Summary}
\label{sec:summary}

In summary, we report the GMRT detection of 21cm absorption from the $z \sim 3.39$ DLA 
towards \pks, with the 21cm profile consisting of two separate components spread over 
$\sim 115$~\kms. The stronger of these components is in excellent agreement in both 
redshift and FWHM with the feature tentatively detected by \citet{briggs97} with the Arecibo 
telescope. We suggest that the higher peak optical depth measured by the latter could be due to
a combination of an incorrect flux density scale and low-level RFI in the single-dish 
spectrum, resulting in an over-estimate of the line depth. We find no significant evidence 
for optical depth variability in the 21cm line.

We obtain a covering factor of $\sim 0.69$ from a 328~MHz VLBA image and use this to
estimate an average spin temperature of $\ts \sim [(955 \pm 160) \times (f/0.69)]$~K, 
assuming that the average column density across the radio core is the same as that measured
from the Lyman-$\alpha$ line. A comparison between the optical low-ionization metal-line 
and 21cm profiles finds evidence for structure in the absorbing gas across the radio core,
bringing the above assumption into question. Despite this, the GMRT observations indicate that 
most of the HI towards the radio core must be in the warm phase, even if a high CNM 
fraction is present on the narrow sightline towards the optical QSO. Finally, the complexity 
of the 21cm  and CII* absorption lines and the large offset between peak optical and 21cm 
redshifts suggests that the $z \sim 3.39$ DLA is unlikely to be of use in probing
fundamental constant evolution.

\section{Acknowledgments}
We thank Sara Ellison, Elias Brinks and Frank Briggs for respectively providing us with the 
Keck-Hires spectrum towards PKS~0201+113, a VLA 327~MHz image of the PKS~0201+113 field and 
a postscript version of the VLA 21cm spectrum, and Ayesha Begum for carrying out the GMRT 
observations of 2003 September. We also thank the staff of 
the GMRT who made these observations possible. The GMRT is run by the National 
Centre for Radio Astrophysics of the Tata Institute of Fundamental Research. Basic 
research in radio astronomy at the Naval Research Laboratory is supported by
6.1 base funding. The National Radio Astronomy Observatory is operated by 
Associated Universities, Inc, under cooperative agreement with the National Science 
Foundation.

\bibliographystyle{mn2e}
\bibliography{ms}

\end{document}